\documentclass[aps,twocolumn,amssymb,amsfonts,amsmath,showpacs,final,a4paper,superscriptaddress]{revtex4-1}

\usepackage{float}
\usepackage[dvips,final]{graphicx}
\usepackage{color}
\usepackage{amsmath}

\newcommand{\RM}[1]{\MakeUppercase{\romannumeral #1{.}}}
\begin{document}

\title{Separation of ultrafast spin currents and spin-flip scattering in Co/Cu(001) driven by femtosecond laser excitation via the complex MOKE}

\author{J. Wieczorek}
\affiliation{Faculty of Physics and Center for Nanointegration
(CENIDE), University of Duisburg-Essen, Lotharstr.~1, 47057
Duisburg, Germany}

\author{A. Eschenlohr}
\affiliation{Faculty of Physics and Center for Nanointegration
(CENIDE), University of Duisburg-Essen, Lotharstr.~1, 47057
Duisburg, Germany}

\author{B. Weidtmann}
\affiliation{Faculty of Physics and Center for Nanointegration
(CENIDE), University of Duisburg-Essen, Lotharstr.~1, 47057
Duisburg, Germany}

\author{M. R\"{o}sner}
\affiliation{Institute for Theoretical Physics, Bremen Center for
Computational Materials Science,University of Bremen,
Otto-Hahn-Allee 1, 28359 Bremen, Germany}

\author{N. Bergeard}
\affiliation{Faculty of Physics and Center for Nanointegration
(CENIDE), University of Duisburg-Essen, Lotharstr.~1, 47057
Duisburg, Germany}

\author{ A. Tarasevitch}
\affiliation{Faculty of Physics and Center for Nanointegration
(CENIDE), University of Duisburg-Essen, Lotharstr.~1, 47057
Duisburg, Germany}

\author{T. O. Wehling}
\affiliation{Institute for Theoretical Physics, Bremen Center for
Computational Materials Science,University of Bremen,
Otto-Hahn-Allee 1, 28359 Bremen, Germany}

\author{U. Bovensiepen}\email[] {uwe.bovensiepen@uni-due.de}
\affiliation{Faculty of Physics and Center for
Nanointegration (CENIDE), University of Duisburg-Essen,
Lotharstr.~1, 47057 Duisburg, Germany}

\date{\today}

\begin{abstract}

Ultrafast magnetization dynamics in metallic heterostructures
consists of a combination of local demagnetization in the
ferromagnetic constituent and spin-dependent transport
contributions within and in between the constituents. Separation
of these local and non-local contributions is essential to obtain
microscopic understanding and for potential applications of the
underlying microscopic processes. By comparing the ultrafast
changes of the polarization rotation and ellipticity in the
magneto-optical Kerr effect (MOKE) we observe a time-dependent
magnetization profile $M(z,t)$ in Co/Cu(001) films by exploiting
the effective depth sensitivity of the method. By analyzing the
spatio-temporal correlation of these profiles we find that on time
scales before hot electron thermalization ($<100$~fs) the
transient magnetization of Co films is governed by spin-dependent
transport effects, while after hot electron thermalization
($>200$~fs) local spin-flip processes dominate.

\end{abstract}



\maketitle

\section{introduction}

Electronic excitations in ferromagnets are essentially spin
polarized. In the transition metal ferromagnets Fe, Co, and Ni
excitations of 0.1--1~eV relax on femto- to picosecond time scales
\cite{aeschlimann1997,goris2011} due to scattering with secondary
excitations mediated by electron-phonon, electron-magnon, and
electron-electron / exchange interaction. The underlying
microscopic processes are essential
 in ultrafast magnetization dynamics \cite{beaurepaire1996,kirilyuk2010}.

Recently, this field has been propelled by ultrafast spin
polarized \cite{malinowski2008,battiato2010,melnikov2011,
rudolf2012,kampfrath2013,graves2013} and unpolarized currents
\cite{eschenlohr2013} generated by a gradient in excitation
density \cite{brorson1987}. These effects offer to exploit
non-local magnetization dynamics. Already now femtosecond (fs)
laser-excited spin currents in layer stacks were reported to
induce spin transfer torque \cite{schellekens2014}. Furthermore,
laser-induced spin currents in a magnetic tunnel junction were
controlled by a bias voltage \cite{savoini2014}. Moreover, spin
currents are reported to drive an ultrafast change between ferri-
and ferromagnetic order \cite{graves2013}.

Such demonstrations suggest an extension of spintronics, including
spin filter and magnetoresistance effects, into the
non-equilibrium regime. In turn, fs time-resolved experiments
allow conclusions on spin transport \cite{malinowski2008,
melnikov2011, rudolf2012, eschenlohr2013, kampfrath2013}, to shed
light on the underlying microscopic mechanisms. However, up to
date experimental means to distinguish ultrafast spin currents,
which propagate into conducting contacts/substrates from
ferromagnetic layers, and competing depolarizing spin-flip
scattering, which reduces the magnetization $M$ in these spin
current sources, are missing.

Here we show how to fill this gap by establishing that
laser-excited spin currents result in a specific transient
magnetization profile $M(z,t)$ in the direction $z$ along the film
normal, before spin flips impact the dynamics. We employ
ferromagnetic Co/Cu(001) films as a model system for a spin
current source contacted by a spin sink. Key is the combined
analysis of the film thickness dependence with the time-dependent
difference in polarization rotation $\theta$ and ellipticity
$\epsilon$ of the magneto-optical Kerr effect (MOKE), which
provides a   depth sensitivity and access to $M(z,t)$. We separate
spin currents generated by non-thermalized carriers up to
$\sim$100~fs and local demagnetization due to spin-flip scattering
of thermalized carriers dominant after $\sim$200~fs.

A contested issue in ultrafast MOKE is whether the observations
reflect $M(t)$, or transient changes of the optical constants.
Koopmans et al. \cite{koopmans2000}, finding that $\theta$ and
$\epsilon$ of the complex Kerr angle $\Phi = \theta +
i\cdot\epsilon$ showed a different behavior for $t\le 100$~fs,
assigned the difference to state-filling effects. Later, Guidoni
et al. \cite{guidoni2002} reported that the magneto-optical
response is dominated by the magnetization dynamics after electron
thermalization. However, the origin of the transient difference
between $\theta$ and $\epsilon$ before thermalization has so far
been unclear \cite{guidoni2002}. We derive that the transient
difference of $\theta$ and $\epsilon$ on femtosecond time scales
is a result of a spatial profile in the magnetization.

Also, the relative importance of (i) spin-flip scattering of
thermalized electrons, described by the microscopic three
temperature model (M3TM) \cite{koopmans2009}, and (ii)
superdiffusive spin transport
\cite{battiato2010,battiato2012,yastremsky2014} of hot,
non-equilibrium electrons, in ultrafast magnetization dynamics is
under discussion. Experiments on layered structures have
demonstrated the importance of spin transport
\cite{malinowski2008, melnikov2011, rudolf2012, eschenlohr2013,
kampfrath2013}, while  demagnetization of metallic films on
insulators was ascribed to spin flips \cite{schellekens2013},
leading to an emerging consensus that both processes play a role
\cite{turgut2013}. Here, we investigate both spin-flip processes
confined to the ferromagnetic film, as well as propagating spin
currents that are accepted by the metallic substrate, in a model
system.

\section{analysis of transient magnetization gradients}

Femtosecond laser excitation of a heterostructure like Co/Cu(001)
epitaxial films, which present a large gradient in spin
polarization from the film to the substrate already in
equilibrium, leads to spatial redistribution of charge carriers
\cite{brorson1987,lisowski2004} and spin polarization by transport
effects in particular across the interface \cite{battiato2010,
melnikov2011,turgut2013}. Locally within the ferromagnetic film
the magnetization changes also by spin-flip scattering. In order
to quantify the spin-dependent transport processes compared to
local demagnetization we first estimate these two contributions by
analytical calculations in one dimension along the interface
normal direction. Lateral magnetization gradients within the film
plane are also present due to the lateral gradient in excitation
density following the laser focus with a full width at half
maximum, in the present case, of about 16~$\mu$m. However, since
these lateral gradients are three orders of magnitude smaller than
the one along the normal direction, we neglect them here and
consider only the normal component, which is essential for the
sub-picosecond dynamics.

\subsection{Estimation of spin diffusion and local demagnetization contributions}

\begin{figure}
    \centering
        \includegraphics[width=0.95\columnwidth]{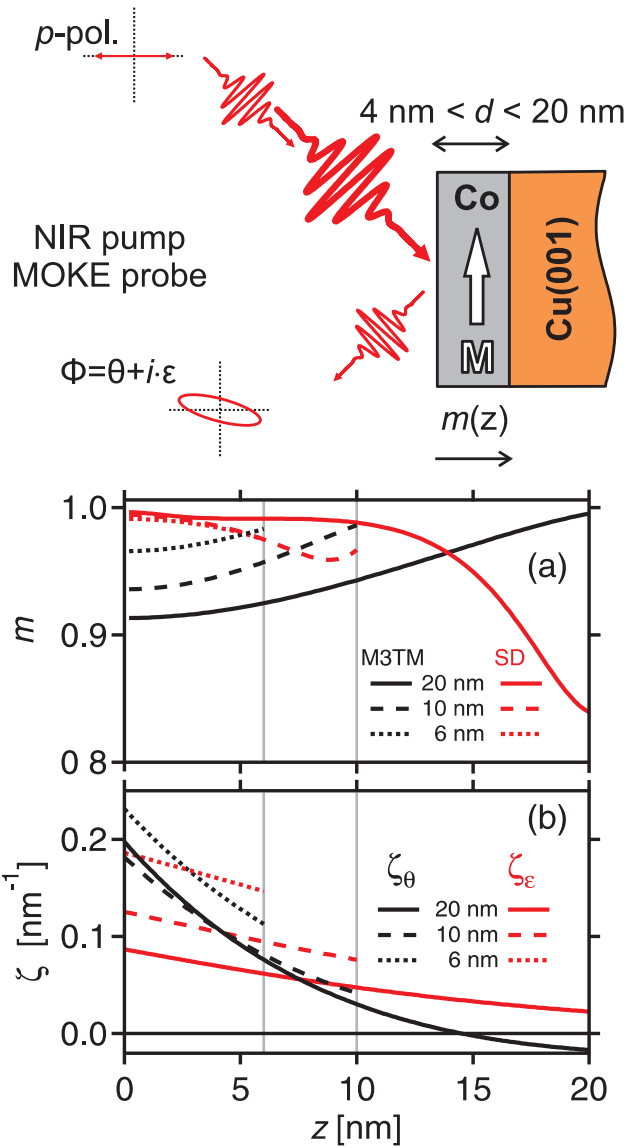}
    \caption{
(Top) Kerr rotation and ellipticity of Co/Cu(001) are measured in
a pump-probe experiment. (a) Normalized magnetization profile
$m(z)$ calculated with the M3TM \cite{koopmans2009} (black) for
$t_0=170$~fs and spin diffusion (SD, red) for $t_0=20$~fs. (b)
MOKE sensitivity $\zeta(z)$ for the real part/rotation
$\zeta_{\theta}(z)$ (black) and imaginary part/ellipticity
$\zeta_{\epsilon}(z)$ (red). The respective curves are shown for
Co thicknesses $d$ of 6 (dotted), 10 (dashed), 20~nm (solid
lines). }
    \label{fig:idee}
\end{figure}

In Fig.~\ref{fig:idee}(a) we contrast the relative local
magnetization $m(z,t)=M(z,t)/M_0$ for (i) spin-flip scattering of
thermalized electrons and (ii) spin transport, calculated with an
extended M3TM and a spin diffusion equation, respectively. $M_0$
is the homogeneous equilibrium magnetization.

To estimate the spin-flip contribution we start with calculating
the time- and spatially dependent electronic and lattice
temperatures, $T_{\mathrm{e}}(z,t)$ and $T_{\mathrm{l}}(z,t)$,
respectively, for different Co film thicknesses $d$ on Cu(001)
employing a two temperature model (2TM) \cite{bonn2000}. We
include electronic heat diffusion, which is driven by the spatial
gradient in the electronic temperature \cite{bonn2000}, and
account for the Co film with thickness $d$. We further include the
Cu substrate in the calculation. We take the optical penetration
depth $\delta_{\mathrm{skin}}$ of the pump pulse at 800~nm central
wave length into account and include thereby the position
dependent optical absorption in the calculation. The initial
excitation profile in cobalt and copper was calculated by
\begin{equation}
S(z,t)=\frac{F_{\mathrm{abs},i}\cdot
e^{-z/\delta_{\mathrm{skin,}i}}}{\delta_{\mathrm{skin,}i}\cdot(1-e^{-d_{i}/\delta_{\mathrm{skin,}i}})}\cdot
G(t)
\end{equation}

$G(t)$ is a normalized Gaussian function with a FWHM of 35~fs,
$d_{i}$ is the film thickness, where $i$ stands for cobalt and
copper. For copper we used $d_{\mathrm{Cu}}=\infty$. The
penetration depth of the laser $\delta_{\mathrm{skin,Co}}$=13~nm
and $\delta_{\mathrm{skin,Cu}}$=12~nm was calculated according to
the optical constants $n_{\mathrm{Co}}=2.56+i\cdot4.92$ and
$n_{\mathrm{Cu}}=0.26+i\cdot5.26$ \cite{Palik}. These optical
constants were also used to calculate the absorbed fluence
$F_{\mathrm{abs,}i}$ employing \cite{imd}, where we set the
incident fluence to
$F_{\mathrm{in}}=3~\textnormal{mJ/}\textnormal{cm}^{2}$.

Subsequently, we used the M3TM, which considers spin-flip
scattering only for a thermalized electronic system, to obtain
$dm_{\mathrm{M3TM}}/dt$ and $m_{\mathrm{M3TM}}(z,t)$
\cite{koopmans2009}. To take the thermalization time of the
electronic system in the simulation for the spin-flip scattering
into account, we multiply the expression $dm_{\mathrm{M3TM}}/dt$
from the M3TM \cite{koopmans2009} with a time dependent factor
$n_{\mathrm{therm}}(t)$ \cite{lisowski2004}. It was calculated
with
$n_{\mathrm{hot}}/dt=G(t)-n_{\mathrm{hot}}/\tau_{\mathrm{therm}}$
and
$dn_{\mathrm{therm}}/dt=n_{\mathrm{hot}}/\tau_{\mathrm{therm}}$,
where $n_{\mathrm{hot}}(t)$ and $n_{\mathrm{therm}}(t)$ are
factors between $0 < n < 1$, $G(t)$ is a normalized Gaussian
function with a FWHM of 35~fs and $\tau_{\mathrm{therm}}=150$~fs
is the thermalization time of the electronic system, which is a
typical time for a 3\textit{d} metal \cite{rhie2003}. The used
parameters are the electron-phonon coupling constants
$g_{\mathrm{Co}}=9.3\cdot10^{7}~\textnormal{W/m}^{3}\cdot\textnormal{K}$,
$g_{\mathrm{Cu}}=10^{7}~\textnormal{W/m}^{3}\cdot\textnormal{K}$
\cite{hohlfeld1998}, the electronic heat capacity coefficients
$\gamma_{\mathrm{Co}}=4.4~\textnormal{mJ/mol}\cdot
\textnormal{K}^{2}$,
$\gamma_{\mathrm{Cu}}=0.69~\textnormal{mJ/mol}\cdot
\textnormal{K}^{2}$ \cite{tari2003}, the specific heats for $T
\rightarrow \infty$
$C_{\mathrm{Co}}=24.81~\textnormal{J}/\textnormal{mol}\cdot\textnormal{K}$,
$C_{\mathrm{Cu}}=24.43~\textnormal{J}/\textnormal{mol}\cdot\textnormal{K}$,
the Debye temperatures $\Theta_{\mathrm{Co}}=386$~K,
$\Theta_{\mathrm{Cu}}=310$~K \cite{ho1974}, the mass densities
$\rho_{\mathrm{Co}}=8.86~\textnormal{g/cm}^{3}$,
$\rho_{\mathrm{Cu}}=8.96~\textnormal{g/cm}^{3}$ \cite{yaws2005},
the molar masses $M_{\mathrm{Co}}=58.93~\textnormal{g/mol}$,
$M_{\mathrm{Cu}}=63.55~\textnormal{g/mol}$ \cite{wieser2011}, the
thermal conductivities
$\kappa_{\mathrm{Co}}=100~\textnormal{W/m}\cdot \textnormal{K}$,
$\kappa_{\mathrm{Cu}}=400~\textnormal{W/m}\cdot \textnormal{K}$
\cite{ho1974}, the Curie temperature of Co $T_{\mathrm{C}}=1388$~K
\cite{stearns1986}, and the M3TM scaling factor for the
demagnetization rate of Co $R=25.3~\textnormal{ps}^{-1}$
\cite{koopmans2009}.

To simulate spin diffusion (SD) we employ a diffusion equation for
excited electrons based on Fick's second law \cite{crank1980} for
majority and minority electrons. Eq.~\ref{Eq:Fickslaw} contains
the respective source, diffusion and decay terms. We discard
charge current contributions because contrary to dielectrics a
potential charge current which is linked to the spin current is
screened in metals already on time scales slower than the inverse
plasma frequency. In Co these time scales were concluded to be
sub-fs \cite{braicovich2008} and are thus well below our
experimental time resolution. We use the diffusion coefficient
$D_{\sigma}=v_{\sigma} \lambda_{\sigma}/3$, $\sigma$ represents
majority and minority electrons, for self diffusion to simulate
the transport. The respective mean free path along $z$ is
$\lambda_{\sigma}=v_{\sigma} \tau_{\sigma}$. Velocities
$v_{\sigma}$ and lifetimes $\tau_{\sigma}$ were averaged up to
0.5~eV above the Fermi level $E_{\mathrm{F}}$ which is reasonable
to account for spin diffusion considering that the experimentally
determined scattering times reported in
Refs.~\cite{aeschlimann1997,goris2011} refer to primary excited
electrons. An electron at an energy of $E-E_{\mathrm{F}}=1$~eV
has, e. g., a lifetime below 10 fs. Within our time resolution,
such a primary excited electron will have scattered more than once
and lowered its energy accordingly. We used energy averaged
lifetimes $\tau_{\uparrow}=22~\textnormal{fs}$ and
$\tau_{\downarrow}=20~\textnormal{fs}$ which were determined from
Ref.~\cite{goris2011}. The velocities were calculated by density
functional calculations (see Appendix) which resulted in energy
averaged velocities $v_{\uparrow}=0.6~\textnormal{nm/fs}$ and
$v_{\downarrow}=0.2~\textnormal{nm/fs}$. Taking these values for
$\lambda$ and $v$ we obtained the considered diffusion constants
$D_{\uparrow}=2.4~\textnormal{m}^{2}/\textnormal{fs},
D_{\downarrow}=0.27~\textnormal{m}^{2}/\textnormal{fs}$. We
finally calculate the spin diffusion by

\begin{equation}\label{Eq:Fickslaw}
\frac{\partial n_{\sigma}}{\partial t}= D_{\sigma}\cdot
\frac{\partial n_{\sigma}^{2}}{\partial z^{2}}+c_{\sigma}\cdot
S_{\sigma}(z,t)-\frac{n_{\sigma}}{\tau_{el}},
\end{equation}

where $n_{\sigma}$ is the density of excited electrons,
$c_{\sigma}S_{\sigma}(z,t)$ describes fs laser excited primary and
subsequently generated secondary electrons with $c_{\sigma}$ being
a constant factor. To account for the spatial distribution, we
convoluted two exponential decay terms according to the position
dependent absorption due to the optical penetration length
$\delta_{skin}$ and the spin dependent mean free path
$\lambda_{\mathrm{ball,}\sigma}$ of primary excited electrons
\cite{hohlfeld2000}. The latter is determined by ballistic
velocities and lifetimes which were taken at energies of the
maximum excitation probability in the joint density of states, see
Appendix. The values are
$v_{\textnormal{ball},\uparrow}=1~\textnormal{nm/fs}$,
$v_{\textnormal{ball},\downarrow}=0.3~\textnormal{nm/fs}$ and
$\tau_{\textnormal{ball},\uparrow}=18~\textnormal{fs}$,
$\tau_{\textnormal{ball}\downarrow}=6~\textnormal{fs}$. As an
effective decay time of the spin polarized current we took a value
of $\tau_{\mathrm{el}}=100$~fs \cite{battiato2012}. Then,
$\lambda_{\mathrm{ball,}\sigma}$ was calculated with the
respective velocities and lifetimes. We included propagation in Cu
with $\lambda=70$~nm \cite{hohlfeld2000} by setting $n_{\sigma}=0$
since bulk Cu acts as an efficient electron and spin drain. The
last term $-n_{\sigma}/\tau_{el}$ describes electron
thermalization and spin current decay by a time constant
$\tau_{\mathrm{el}}$. The change $\partial
m_{\mathrm{SD}}/\partial t$ is defined through the balance of the
magnetic moments of in- and outgoing electrons.

Comparing the magnetization profiles resulting from local
spin-flip scattering and spin transport, as depicted in
Fig.~\ref{fig:idee}(a), $m(z,t_0)$ are strikingly and
systematically different for the two scenarios because the
respective gradients $\partial m(z,t_0) / \partial z$ have an
opposite sign. While consideration of of spin-flip scattering of
thermalized electrons results in a minimum $m$ at the surface and
a maximum at the Co-Cu interface, the spin diffusion description
leads to a depletion of $m$ at the interface in combination with a
weak variation near the surface, in good agreement with a more
sophisticated description by superdiffusive spin transport
\cite{battiato2012}.

\subsection{Measurement of the spatio-temporal magnetization dynamics}

To analyze $m(z,t)$ experimentally, we measured the complex MOKE
$\Phi=\theta+i\cdot\epsilon$ in a pump-probe experiment, see
Fig.~\ref{fig:idee}(top). The magneto-optical (MO) Kerr rotation
$\theta$ and  the ellipticity $\epsilon$ were determined by a
polarization analysis using a balance detection scheme. We
employed a cavity dumped Ti:sapphire oscillator which generates
p-polarized 35~fs laser pulses at $h\nu=1.55$~eV and 40~nJ pulse
energy, that are split into pump and probe pulses at a 4:1 ratio,
at 2.53~MHz repetition rate. The incident pump fluence was
6~mJ/cm$^2$. Further details of the experimental setup are given
in Ref.~\cite{sultan2011, sultan2012}. Epitaxial Co films of
4~nm~$<d<$~20~nm, which we investigated \textit{in situ}, were
grown in ultrahigh vacuum on Cu(001) following Ref.
\cite{berger1992}. The film's easy axis of the magnetization lies
in the film plane and we measure MOKE in the longitudinal
geometry.


\begin{figure}
    \centering
        \includegraphics[width=0.95\columnwidth]{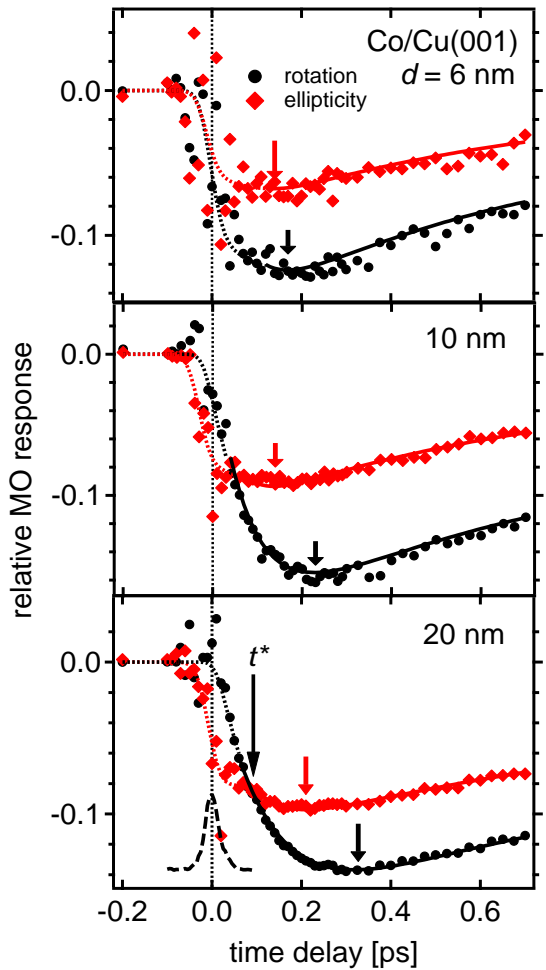}
\caption{Time-dependent relative MO responses of Kerr rotation
$\Delta \theta(t)/\theta_{0} $ (dots) and ellipticity $\Delta
\epsilon(t)/\epsilon_{0} $ (diamonds) of Co/Cu(001) for $d=6$, 10,
20~nm. Solid lines represent analytical fit functions, while dotted lines guide
the eye. For determination of $t=0$ we measured the auto
correlation function with second harmonic generation from the
surface (dashed line, bottom panel). Arrows indicate the delay of minimum MO
response; $t^\star$ indicates when $\Delta \theta(t)/\theta_{0}
=\Delta \epsilon(t)/\epsilon_{0} $.}
    \label{fig:data}
\end{figure}

We find different transient responses of $\theta(t)$ and
$\epsilon(t)$, shown in Fig. \ref{fig:data}. Our following
conclusions are based on the fact that Kerr rotation $\theta$ and
ellipticity $\epsilon$ exhibit different depth sensitivities which
we  use to identify specific magnetization profiles $m(z,t)$
caused by spin-flip scattering of thermalized electrons or spin
polarized currents, see Fig.~\ref{fig:idee}(a). In
Fig.~\ref{fig:idee}(b) we plot
$\zeta=\zeta_{\theta}+i\cdot\zeta_{\epsilon}$, which we term MOKE
sensitivity for the complex Kerr angle
$\Phi=\theta+i\cdot\epsilon$, as a function of the position $z$
within a 6, 10, and 20~nm thick film. The sensitivity $\zeta$
indicates how changes $\Delta m(z,t)=m(z,t)-1$ contribute at a
depth $z$ in first order to the total change. The measured
transient changes in $\Phi$ are calculated by integration over the
Co film thickness

\begin{equation}\label{Equation:zeta}
\Delta \theta (t) / \theta_{0} + i \cdot \Delta \epsilon (t) /
\epsilon_{0} = \int_{0}^{d} \zeta(z)\cdot\Delta m(z,t) dz.
\end{equation}

Following Traeger et al. \cite{traeger1992}, $\zeta(z)$ was
calculated by setting $m=0$ with the exception of a part $z,z+dz$
where $m=1$, and determining the complex Kerr angle of this part
$\Phi_{dz}=\theta_{dz}+i\epsilon_{dz}$ for this system. To take
pump-induced changes into account we normalized $\Phi_{dz}$ to the
equilibrium values and obtain

\begin{equation}
\zeta(z)\cdot
dz=\frac{\theta_{dz}}{\theta_{0}}+i\cdot\frac{\epsilon_{dz}}{\epsilon_{0}}.
\end{equation}

We used the matrix formalism by Zak et al. \cite{zak1991} with
$dz=0.2$~nm and refractive indices $n_{\mathrm{Cu}}$ and
$n_{\mathrm{Co}}$. The  static MO constant $q_{\mathrm{Co}}$ which
enters the determination of $\zeta$ was determined experimentally
to $q_{\mathrm{Co}}=0.017-i\cdot 0.020$ by fitting the thickness
dependent MO contrast and the ratio $\epsilon/\theta$ for
$d=2-20$~nm (not shown).

Note that $\zeta_{\theta}$ is larger near the surface than near
the Co-Cu interface and $\zeta_{\epsilon}$ exhibits a weaker
$z$-dependence, see Fig.~\ref{fig:idee}(b). Applying  eq.
(\ref{Equation:zeta}) reveals that (i) $\theta$ probes effectively
the near surface part. For thicker layers like 20~nm the sign of
the rotation sensitivity eliminates some signal contribution from
the bulk part of the film which also leads to an effective probing
of the surface near region. (ii) The second magneto-optical
observable $\epsilon$ rather averages over the film and includes a
sensitivity at the Co/Cu interface. This depth dependence of
$\zeta$ provides us with a probe for $m(z,t)$.


Now we consider our experimental results in detail.
Fig.~\ref{fig:data} depicts the time dependent relative MO
responses $\Delta \theta(t)/\theta_{0}$ and $\Delta
\epsilon(t)/\epsilon_{0}$ for Co/Cu(001) for $d = 6, 10, 20$~nm.
The curves show a reduction of the signal which begins within the
laser pulse and a subsequent recovery starting between 100 and 300
fs, depending on the film thickness. Most importantly, we find for
all $d$ differences for $\theta (t)$ and $\epsilon (t)$. For $d
\geq 10$~nm we observe a faster reduction of $\epsilon$ than for
$\theta$ and a crossing at delay $t^{\star}$ of both curves within
200~fs. After this crossing the changes are inverted relative to
each other and $\Delta \theta(t)/\theta_{0}$ remains stronger than
$\Delta \epsilon(t)/\epsilon_{0}$ up to 25~ps. This remaining
difference will be addressed further below. For $d<10$~nm the
initially stronger reduction of $\epsilon$ compared to $\theta$ is
not observed, likely due to a limited pulse duration and a
remainder of the coherent artifact \cite{radu2009}, in combination
with a weaker $dm/dz$. Changes for such $d$ are larger in $\theta$
than in $\epsilon$.


\begin{figure}
    \centering
        \includegraphics[width=0.95\columnwidth]{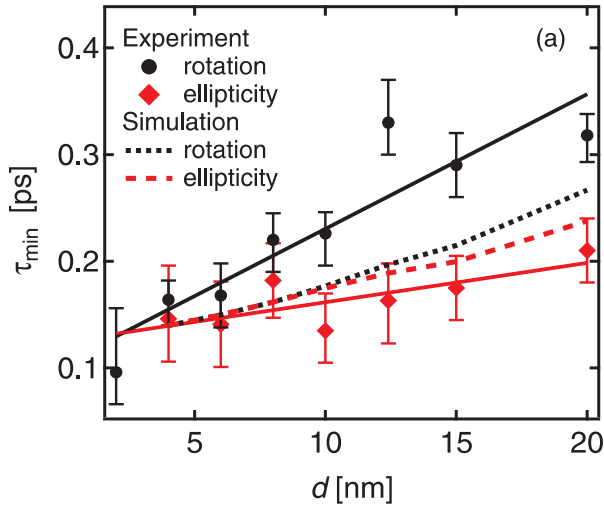}
\caption{(a) Time delay $\tau_{\textnormal{min}}$ of minimal
response in $\Delta \theta(t)/\theta_{0} $ (black dots) and
$\Delta \epsilon(t)/\epsilon_{0} $ (red diamonds) vs. $d$. Solid
lines are linear fits, dashed lines depict simulations.}
    \label{fig:minima}
\end{figure}

From the observed $\theta(t)$ and $\epsilon(t)$ we conclude on
$m(z,t)$ and relate this to a spin polarized current or spin-flip
scattering, see Fig.~\ref{fig:idee}(a). Here, the different depth
sensitivity of $\theta$ and $\epsilon$, see
Fig.~\ref{fig:idee}(b), is essential. Remember, $\epsilon$
exhibits a stronger sensitivity at the Co/Cu interface than
$\theta$. The stronger reduction of the ellipticity compared to
the rotation starting within the pulse duration, see
Fig.~\ref{fig:data}, thus implies that the film is demagnetized
more near the inner interface than near the surface. The
respective magnetization profile agrees qualitatively with the one
obtained for SD, see Fig.~\ref{fig:idee}(a). After some tens of
femtoseconds, depending on $d$, $m(z,t)$ is concluded to be
redistributed towards a profile expected from the M3TM
\cite{koopmans2009}, as can be seen in the larger variation of
$\theta$ compared to $\epsilon$ in Fig.~\ref{fig:data} at
$t>t^\star$, taking into account that $\theta$ has a higher
sensitivity at the surface than $\epsilon$, see
Fig.~\ref{fig:idee}(b). These transient changes of the
magnetization profile suggest that when the electronic system has
not yet thermalized, a spin polarized current dominates $m(z,t)$.
With electron thermalization, spin-flip events take over. This is
consistent with both the superdiffusive transport model
\cite{battiato2012} and the M3TM \cite{koopmans2009}. The
superdiffusive transport model predicts spin transport for a
non-thermalized electron system, which recedes with thermalization
\cite{battiato2013}, while the M3TM considers only thermalized
electrons, which contribute to spin-flip scattering
\cite{koopmans2009}.

Our conclusions are corroborated by finding that the time delay
$\tau_{\textnormal{min}}$ of the minimum MO response exhibits a
systematic thickness dependence, as depicted in
Fig.~\ref{fig:data}. To determine this delay of the minimum
response we performed a fitting analysis and determined
$\tau_{\textnormal{min}}$ using
$f(t)=A_{M}\cdot(1-\exp(-t/\tau_{M}))+A_{R1}\cdot(1-\exp(-t/\tau_{R1}))+A_{R2}\cdot(1-\exp(-t/\tau_{R2}))$;
$A$ and $\tau$ are the amplitudes and time constants,
respectively, $M,R1,R2$ refer to demagnetization and recovery
\cite{note_pulsedurationbroadening}. For $\epsilon$ and $\theta$
the minimum positions shift to later times with increasing $d$,
see Fig.~\ref{fig:minima}. Particularly interesting is the
different thickness dependence of $\epsilon$ and $\theta$. By
fitting the minimum's position as a function of $d$ with a line,
we find with slopes of $b_{\theta}= 13$~fs/nm and $b_{\epsilon}=
4$~fs/nm a pronounced, three time difference.

\begin{figure}
    \centering
        \includegraphics[width=0.95\columnwidth]{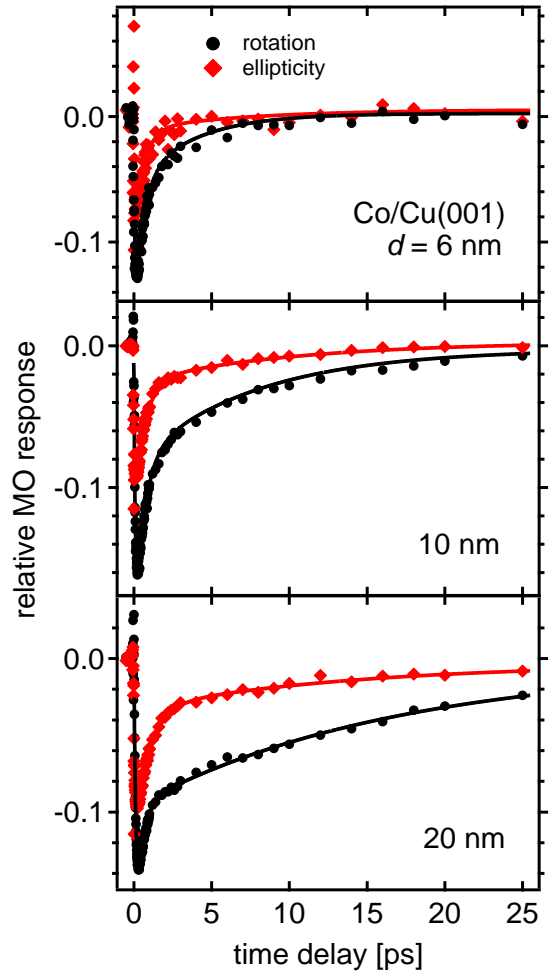}
\caption{Time-dependent relative magneto-optical responses of Kerr
rotation $\Delta \theta(t)/\theta_{0} $ (dots) and ellipticity
$\Delta \epsilon(t)/\epsilon_{0} $ (diamonds) of Co/Cu(001) for
$d=6$, 10, 20~nm for pump-probe delays of up to 25~ps. Solid lines
represent analytical fits described in the text. }
    \label{fig:long}

\end{figure}

Fig.~\ref{fig:long} shows time-dependent MOKE data for longer time
delays up to 25~ps for selected film thicknesses. The difference
between the MOKE ellipticity and rotation remains clearly longer
than 5~ps and the time delay of merging of the two observables
shifts to later delays with increasing film thickness. Considering
typical electronic relaxation or electron-lattice equilibration
times of few ps suggests that this difference cannot be fully
explained by the competition of spin diffusion and local
demagnetization. We will come back to this aspect in the
discussion section.

\subsection{Model description of the spatio-temporal magnetization dynamics}

We simulated the thickness dependent $m(z,t)$ with the extended
M3TM \cite{koopmans2009} and the spin polarized diffusion
equation, as introduced above, by adding both contributions

\begin{equation}
\frac{dm(z,t)}{dt}=\frac{\partial m_{M3TM}(z,t)}{\partial
t}+\frac{\partial m_{SD}(z,t)}{\partial t}.
\end{equation}

\begin{figure}
    \centering
        \includegraphics[width=0.95\columnwidth]{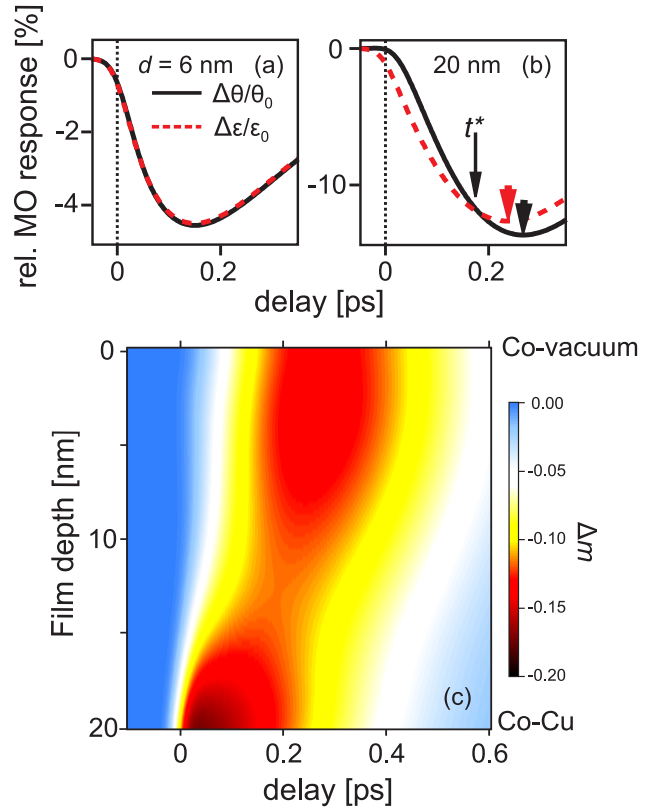}
\caption{(a,b) Simulated magneto-optical response $\Delta
\theta(t)/\theta_{0} $ (solid) and $\Delta
\epsilon(t)/\epsilon_{0} $ (dashed) for $d=6,20$~nm. (c) Simulated
spatio-temporal variation of the relative magnetization change
$\Delta m$ for a 20~nm thick Co film on Cu(001) in a false color
representations as a function of time delay and position within
the film.}
    \label{fig:simulation}
    \end{figure}

We calculated the corresponding MO response with
eq.~\ref{Equation:zeta} using the parameters given above in Sec.
II.A, and determined the delays of minimum MO response as a
function of film thickness $d$, which are included in
Fig.~\ref{fig:minima} as dashed lines. The overall shift of
$\tau_{min}$ to later times with increasing $d$ is explained by
the lower heat conductivity of Co compared to Cu \cite{ho1974}.
Essential for the discussion of the competing spin-dependent
processes is the thickness-dependent difference between
$\tau_{\textnormal{min}}$ for $\theta$ and $\epsilon$, which is
qualitatively reproduced by our simulation. The Co-Cu interface
region, which is preferentially probed by $\epsilon$, reaches the
minimum magnetization earlier and also starts to recover earlier.
Consequently, this confirms our above explanation of the initial
dynamics by spin transport near the Co/Cu interface and the later
dynamics by spin-flip scattering at the surface. We qualitatively
reproduce the transient behavior of our MO observables already
with such a relatively simple model up to delays of 400~fs when
the electronic system is in non-equilibrium with respect to the
lattice. In Fig.~\ref{fig:simulation}(a,b) we show two simulations
of $\Delta \theta$ and $\Delta \epsilon$ for $d = 6, 20$~nm. For
20~nm we obtain an initially faster ($<100$~fs) reduction of
$\epsilon$ compared to $\theta$, and later at 200~fs the crossing
of both curves. For 6~nm our simulation does not exhibit any
difference for the two magneto-optical observables which is in
agreement with the experimental result as far as for these small
thicknesses the minima in $\Delta \epsilon(t)/\epsilon_0$ and
$\Delta \theta(t)/\theta_0$ are not discernible within the error
bars. We will come back to this point in the discussion below.

Having explained the origin of the different transient behavior of
the two MO observables by the two separate, elementary
spin-dependent processes of spin transport from the Co film into
the substrate and local spin-flip scattering within the Co film,
we simulated the spatio-temporal dynamics of the magnetization
change $\Delta m$ as a function of position within the Co film and
time delay. The result is shown in Fig.~\ref{fig:simulation}(c)
and highlights the separate locations of the spin-flip and the
spin transport processes in the near surface region at delays just
above 200~fs and near the Co-Cu interface earlier than 100~fs.

Note that we used a reduced fluence in the calculation compared to
the experiment in order to adjust the calculated magnitude of
demagnetization to the experimentally observed one. The relative
difference between the fluence in the calculation and in the
experiment is roughly two times which is attributed to deviations
of the optical constants in the epitaxial film-substrate system,
limitations in the determination of the laser focus on the sample
surface in the vacuum chamber, and the employed simplifications in
the model.


\section{discussion}

The calculated behavior with spin-flip and spin current
contributions agrees with the experimental observation, c.f.
Figs.~\ref{fig:data},\ref{fig:minima}. For $t>t^\star$, the pump
induced variation for $\theta$ is stronger than for $\epsilon$,
also in agreement with Fig.~\ref{fig:data}, and the dynamics are
dominated by spin-flip scattering. This behavior is only obtained
if the SD contribution is included, the M3TM alone is not
sufficient. We conclude furthermore that the contributions from
both mechanisms are comparable and therefore both have to be
considered. However, as we demonstrate here, they can be separated
in the time domain. For $d=6$~nm we lose the sensitivity to the
competing processes in the simulation, probably due to a more
homogeneously demagnetized film. After all, Fig.~\ref{fig:idee}(a)
indicates a loss of $m$ due to SD at the inner interface as well
as spin-flip scattering at the surface both on a length scale
comparable to 6~nm. In this thin film limit the magnetization
gradients resulting from spin-flip scattering and spin transport
in the Co film have essentially receded. They become more similar
to each other because the film thickness approaches the
spin-dependent mean free path, as spins can now be transported
into Cu also from regions closer to the Co surface.

The deviation of the simulation and the experimental data in Fig.
\ref{fig:minima} is potentially a result of uncertainties in the
optical constants. In addition, spin-dependent back-diffusion of
electrons from Cu into Co \cite{eschenlohr2013, battiato2012},
which is not included in the simulation, can enhance the
demagnetization near the Co/Cu interface found in our SD
simulations further. Spin-dependent scattering at defects could
also contribute to the spin dynamics and would be determined by
the defect density at the interface. Here, we investigate an
epitaxial film on a single crystal substrate which is the best way
to avoid such defects. As the results of our simulation only show
a qualitative agreement with our experimental results, see
Fig.~\ref{fig:minima}, such approximations might be one of the
reasons for this. For a quantitative description energy and spin
dependent transmission coefficients at the interface need to be
considered, as well as many body renormalizations. This is,
however, beyond the scope of this article. Likely, these aspects
need to be taken into account for the deviation of the fluence
used in the simulation and in the experiment.

A pump-induced variation of the optical constants due to the hot
electron distribution could influence our results, but we are
convinced that $m(z,t)$ dominates the dynamics. We argue as
follows. (i) For films $d\geq 10$~nm the difference between
$\theta$ and $\epsilon$ up to $t=$ 200~fs is comparable to later
times, but with opposite sign (Fig. \ref{fig:data}). To explain
this by a time dependent change of optical constants, a change in
sign at a remaining absolute value would be required. This is
rather unlikely since such effects would decay monotonously with
the hot electron distribution within $\sim1$~ps
\cite{guidoni2002,rhie2003}. (ii) With decreasing $d$ the
difference between $\theta$ and $\epsilon$ at $t<t^\star$ shrinks
faster than for $t>t^\star$ (Fig.~\ref{fig:data}), while the hot
electron distribution remains for all $d$. In contrast, spin
polarized transport is strongly affected by the sample thickness
\cite{melnikov2011,battiato2012,eschenlohr2013}. (iii) The
difference between $\theta$ and $\epsilon$ remains up to 25~ps,
see Fig.~\ref{fig:long}, which is too long for state filling
effects due to a hot electron distribution, considered previously
\cite{koopmans2000}. A magnetization profile $m(z,t)$, which is
detected through the different $\zeta$ for $\theta$ and
$\epsilon$, can, however, remain for such a long time due to a
spatial gradient in the lattice temperature. In fact, the gradient
in the magnetization might persist even longer than it takes the
lattice temperature to homogenize, as at longer timescales, when
the excitation of the electronic system has already relaxed,
changes in the magnetization due to changes in the lattice
temperature are mediated by spin-lattice coupling, which is not
included in the M3TM and has characteristic timescales of several
ps to tens of ps.

\section{conclusion and outlook}

We showed that the different depth sensitivity of the
magneto-optical Kerr rotation and ellipticity can be used to
identify spatial magnetization profiles on ultrafast time scales
resulting from spin polarized transport or spin-flip scattering.
We found that the laser-excited spin dynamics in Co/Cu(001) films
are dominated by spin transport effects on times up to $\sim
100$~fs when the electronic system has not yet thermalized, and by
spin-flip scattering of thermalized electrons subsequently.

Since the sensitivity function, which governs the depth
sensitivity, is straight forward to calculate, the demonstrated
method is readily applicable to further material systems. Here,
analysis of metallic ferromagnets on an insulating substrate might
provide a film substrate combination complementary to the
metal-metal case discussed here. We note that for the here
presented methodology it will be essential to determine the
complex sensitivity function independently of other heterosystems
by a full, thickness dependent analysis, because it is set by the
film-substrate combination rather than by the film alone.
Moreover, an improved experimental interface sensitivity, which
can be obtained by using non-linear optical techniques, promises
direct access to the spin transfer dynamics across interfaces and
might be rewarding in order to obtain deeper insight towards a
full reconstruction of $m(z,t)$ and a more quantitative
understanding.

\begin{acknowledgments}   We are grateful for fruitful discussions
with and experimental support by A. Melnikov. This work was funded
by the DFG through SFB 616, the BMBF through 05K10PG2 FEMTOSPEX,
and the Mercator Research Center Ruhr through Grant No.
PR-2011-0003. The numerical computations were carried out on the
North-German Supercomputing Alliance (HLRN) cluster.
\end{acknowledgments}

\section*{Appendix}

The group velocities as well as the densities of excited carriers
in bulk Co were calculated within the spin-polarized generalized
gradient approximation (GGA) \cite{perdew1996} to density
functional theory as implemented in the Vienna Ab Initio
simulation package (VASP) \cite{kresse1994} with the projector
augmented waves basis sets \cite{boechl1994,kresse1999}. The
calculations were performed with a hcp unit cell of Co (lattice
constants $a \approx 2.51\,$\AA\ and $c \approx 4.07\,$\AA\:
\cite{vincent1967}) using $40 \times 40 \times 40$ k-meshes for
the Brillouin zone integration and a plane-wave cut-off of
$335\,$eV.

The excitation probability for an electron-hole pair formed by
states $|c,k,\sigma\rangle$ (electron) and $|v,k,\sigma\rangle$
(hole) at crystal momentum $k$ and with spin $\sigma$ is
calculated based on Fermi's golden rule according to
  \begin{equation*}
  P(k,\sigma,c,v)=|M^k_{c,v,\sigma}|_{xx}^2 \cdot \delta(E_{c,k,\sigma}-E_{v,k,\sigma} - E_{\rm ex}),
  \label{eq:ex_prob}
  \end{equation*}

where $|M^k_{c,v,\sigma}|_{xx}$ are the optical transition matrix
elements for electric fields polarized in $x$-direction,
$E_{v/c,k}$ conduction and valence band energies, and $E_{\rm
ex}=1.55$~eV is the laser excitation energy. We employed
$0.01\,$eV Gaussian broadening for the evaluation of the
excitation probabilities.

\begin{figure}
\includegraphics[width=0.95\columnwidth]{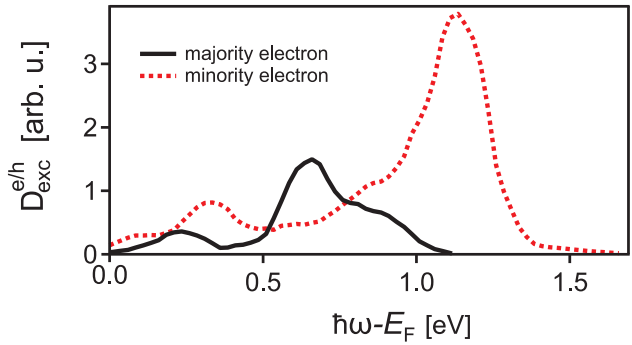}
\caption{Calculated density of carriers excited in Co vs. their
energy with respect to the Fermi level.}
    \label{fig:dft}
\end{figure}

The energy resolved density of excited carriers D is determined by

  \begin{equation*}
    \text{D}_\text{exc}^\text{e/h}(\omega) =
    \frac{1}{N_k}\sum_{c,v,k} |M_k|_{xx}^2\cdot \delta(E^{c,k}_{\sigma}-E^{v,k}_{\sigma} - E_{\rm ex})  \cdot \delta(E^{v/c,k}_{\sigma} - \omega)
        \label{eq:ex_dos}
  \end{equation*}

where $N_k$ is the number of $k$-points. Here, $v$ and $c$ are
band indices referring to states below and above the Fermi level
respectively. The results is depicted in Fig.~\ref{fig:dft}. Using
the above equation we estimate the energy of primary excited
electrons, which yields 0.7~eV and 1.1~eV for the majority and
minority electrons, respectively. The components of the group
velocities perpendicular to the surface, which enter our above
simulation, are calculated according to
$v_{\sigma}(E_{\sigma})=\hbar^{-1}\partial/E_{\sigma}\partial
k_{z} $.

\begin{thebibliography}{99}
\bibitem{note_pulsedurationbroadening}
For the  transient response we multiply this with a Gaussian
function with FWHM 35~fs according to the laser pulse duration of
our experiment.
\end{thebibliography}
\end{document}